**Title: Reconstruction of Bloch wavefunctions of holes in a semiconductor**

**Authors:** J. B. Costello[1]*, S. D. O'Hara[1]*, Q. Wu[1]*, D. C. Valovcin[2], L. N. Pfeiffer[3], K. W. West[3], and M. S. Sherwin[1]†

**Affiliations:**

[1]Physics Department and Institute for Terahertz Science and Technology, University of California, Santa Barbara, CA 93106.
[2] Mathworks, 1 Apple Hill Drive, Natick, MA 01760.
[3]Department of Electrical Engineering, Princeton University, Princeton, NJ 08544.
*These authors contributed equally to this work.

[†]e-mail: sherwin@ucsb.edu

**A central goal of condensed matter physics is to understand how the diverse electronic and optical properties of crystalline materials emerges from the wavelike motion of electrons through periodically-arranged atoms. However, more than 90 years after Bloch derived the functional forms of electronic waves in crystals[1] (now known as Bloch wavefunctions) rapid scattering processes have so far prevented their direct experimental reconstruction. In high-order sideband generation[2-9], electrons and holes generated in semiconductors by a near-infrared (NIR) laser are accelerated to high kinetic energy by a strong terahertz field, and recollide to emit NIR sidebands before they are scattered. Here we reconstruct the Bloch wavefunctions of two types of holes in gallium arsenide at wavelengths much longer than the spacing between atoms by experimentally measuring sideband polarizations and introducing an elegant theory that ties those polarizations to quantum interference between different recollision pathways. These Bloch wavefunctions are compactly visualized on the surface of a sphere. High-order sideband generation (HSG) can, in principle, be observed from any direct-gap semiconductor or insulator. We thus expect the method introduced in this Article can be used to reconstruct low-energy Bloch wavefunctions in many of these materials, enabling important new insights into the origin and engineering of the electronic and optical properties of condensed matter.**

Bloch's theorem tells us how to calculate both the Bloch wavefunctions and the spectrum of electronic energies (the "band structure")[1]. A typical Bloch wave contains spatial oscillations with wavelengths ranging from atomic to macroscopic length scales. Of special interest is the low-energy, long wavelength physics where the excited Bloch waves can be described by a finite-dimensional effective Hamiltonian (Methods). Knowledge about both the low-energy Bloch wavefunctions and band structure are essential to calculating the response of crystalline solids to most external stimuli. The band structure of many crystalline materials can be experimentally reconstructed from angle-resolved photoemission spectroscopy (ARPES) of electrons emitted from their surfaces[10,11]. ARPES enables determination of the energies of electronic waves as functions of their wavelengths and directions of propagation. However, there are no comparably direct methods to reconstruct Bloch wavefunctions. As a result, estimates of Bloch wavefunctions typically depend on parameters derived from fits of complex models[12,13] to a few pieces of experimental data—such as the orbital frequency of an electron in a magnetic field (cyclotron resonance)[14]—that are sensitive only to averages over a range of electronic wave propagation directions and wavelengths. A key obstacle to directly probing Bloch wavefunctions in solids has been that, unlike molecules, where reconstructed electron wavefunctions have been reported[15-17], electronic waves in solids are typically distorted in a few picoseconds by scattering.

Recently, strong laser fields have been used to significantly accelerate electronic waves in solids before they are scattered. For example, high-harmonic generation (HHG) has been demonstrated in solids[18-21] and led to an alternative method to probe band structures[22,23]. However, complicated interference between quantum pathways of electronic waves across multiple bands in HHG experiments[19] hinders the reconstruction of Bloch wavefunctions.

**High-order sideband generation**

In this paper, we present a direct experimental reconstruction of Bloch wavefunctions of holes in bulk gallium arsenide (GaAs) using high-order sideband generation (HSG)[2-9]. In HSG, a relatively weak near-infrared (NIR) laser with frequency $f_{\text{NIR}}$ and a strong laser with terahertz (THz) frequency $f_{\text{THz}}$ simultaneously interact with a semiconductor, resulting in the emission of sideband photons with frequencies $f_{\text{SB}} = f_{\text{NIR}} + n \cdot f_{\text{THz}}$ where $n$ is an integer. If the band structure is symmetric under inversion, as in the (001) plane of GaAs studied here, $n$ must be even.

HSG experiments have been conducted both with extremely narrow-band quasi-cw fields with $f_{THz} < 1$ THz, which have enabled the resolution of sidebands with $n > 60$[7], and broader-band pulsed fields with $f_{THz} > 10$ THz, which have enabled time-resolution of the recollision process even in materials with dephasing times <10 fs[5]. A quasi-cw HSG spectrum from bulk GaAs at 60 K is shown in Fig. 1a. In the experiment, a 100 mW NIR laser, and 2.01±0.13 mJ, 40 nanosecond, 0.447 ± 0.001 THz pulses generated by UCSB's MM-Wave Free-Electron Laser[24] were linearly polarized and collinearly focused on a 500 nm GaAs epilayer (see Fig. 1a inset). The THz electric field strength in the epilayer was 70±2 kV/cm (Methods). In GaAs, HSG can be described by the following three-step process (Fig. 1b, Methods). First, electrons (E) and two species of holes—light holes (LH) and heavy holes (HH)—are created by the NIR laser. Second, the E-LH and E-HH pairs are driven apart and then back towards each other along the direction defined by the THz field (Fig. 1b). Crucially, during this acceleration phase, the Bloch waves associated with the E-LH and E-HH pairs interfere with each other. Third, they recollide with significant kinetic energy and emit sideband photons.

Information about the Bloch wavefunctions sampled by electrons and holes on their journeys through the Brillouin zone is imprinted on the polarization state of each sideband, which we measured by Stokes polarimetry (Methods). Because light from a particular sideband can be associated with a quasi-momentum that is controlled by the THz field, its polarization will be different from the incoherent light emitted at the same energy in the absence of a strong THz field (photoluminescence), which is a superposition of emission from electron-hole pairs with all quasimomenta satisfying energy conservation. The linear orientation angle, $\alpha$, and the ellipticity angle, $\gamma$, for each sideband are shown in Fig. 1c for four different NIR polarizations. The polarizations of sidebands depend on the sideband index $n$ and the NIR polarization in a manifestation of dynamical birefringence[6]. Although sideband intensities have a highly nonlinear dependence on THz power, they are proportional to the NIR power if it is sufficiently small[2]. All data reported here were taken in this regime of linear NIR response. In this linear regime, the sideband polarization can be mapped onto the polarization state of the NIR laser by a dynamical Jones matrix $T$[6], defined in a basis of circularly-polarized fields $\sigma^\pm$ (with helicity ± 1) as

$$\begin{pmatrix} E_{+,n} \\ E_{-,n} \end{pmatrix} = \begin{pmatrix} T_{++,n} & T_{+-,n} \\ T_{-+,n} & T_{--,n} \end{pmatrix} \begin{pmatrix} E_{+,NIR} \\ E_{-,NIR} \end{pmatrix},$$

where $E_{\pm,n}$ and $E_{\pm,NIR}$ respectively denote the $\sigma^\pm$ components of the electric field associated with the $n$-th sideband and NIR laser, and $T_{\pm\pm,n}$ denote the dynamical Jones matrix elements associated with the $n$-th sideband. $T$-matrix elements were determined by measuring the sideband polarizations for four different linear NIR laser polarizations.

**Dynamical Jones Matrices**

To understand the physics underlying each $T$-matrix element, it is necessary to consider the spins of electrons and holes. The four recollision pathways from the excitations generated by the $\sigma_{NIR}^-$ component of the NIR laser are shown in Fig. 2. The electrons have spin 1/2, while the heavy holes (HHs) and light holes (LHs) have total spin 3/2. Driven by the THz field, an electron-hole pair acquires a dynamic phase

$$A_{HH(LH)}(t',t) = -\int_{t'}^{t} dt''(E_c[\mathbf{k}(t'')] - E_{HH(LH)}[\mathbf{k}(t'')])/\hbar,$$

where $E_c$ and $E_{HH(LH)}$ are the energies associated with the E and HH (LH) bands shown schematically in Fig. 1b, $\mathbf{k}$ is the quasimomentum, and $\hbar$ is the reduced Planck's constant (Methods, Supplementary Discussion). The spin ±1/2 of the electron does not change during acceleration. Because the Bloch wavefunctions in both HH and LH bands are superpositions of states with spin $\pm 1/2$ and $\pm 3/2$, the $\sigma^-_{NIR}$ component can generate sidebands with either $\sigma^-_{HSG}$ or $\sigma^+_{HSG}$ while satisfying angular-momentum conservation, giving rise to dynamical Jones matrix elements $T_{--,n}$ and $T_{+-,n}$ respectively. Similar recollision pathways follow from the excitations generated by the $\sigma^+_{NIR}$ component, giving rise to $T_{-+,n}$ and $T_{++,n}$ (Supplementary Discussion, Extended Data Fig. 4).

Properties of dynamical Jones matrices can be derived from the Luttinger Hamiltonian[25], which describes the physics of the HH and LH states at the relatively small energies and quasimomenta probed in this experiment. We tune the NIR laser just below the band gap and direct the THz field to propagate along the $z$-axis to ensure the electrons and holes have no $z$-component of quasimomentum $\boldsymbol{k}$. In this case the Luttinger Hamiltonian takes a block diagonal form

$$H^\pm_v(\boldsymbol{k}) = -\frac{\hbar^2 k^2}{2m_0}[\gamma_1 \tau_0 - 2\gamma_2 \boldsymbol{n}_\pm \cdot \boldsymbol{\tau}] =$$
$$-\frac{\hbar^2 k^2}{2m_0}\begin{pmatrix} \gamma_1 + \gamma_2 & -\sqrt{3}(\gamma_2 \sin(2\theta) \pm i\gamma_3 \cos(2\theta)) \\ -\sqrt{3}(\gamma_2 \sin(2\theta) \mp i\gamma_3 \cos(2\theta)) & \gamma_1 - \gamma_2 \end{pmatrix} \quad \text{(Eq. 1)}$$

where $\tau_0$ is the identity matrix, $\boldsymbol{\tau}$ is the vector of Pauli matrices, $\theta$ is the angle between the THz field and the [110] crystal direction (Fig. 3d inset), $\gamma_1$, $\gamma_2$, and $\gamma_3$ are the scalar Luttinger parameters, $m_0$ is the electron rest mass, and $\boldsymbol{n}_\pm$ is

$$\boldsymbol{n}_\pm = \left(\frac{\sqrt{3}}{2}\sin(2\theta), \mp\frac{\sqrt{3}\gamma_3}{2\gamma_2}\cos(2\theta), -\frac{1}{2}\right) \quad \text{(Eq. 2)}$$

The Bloch wavefunctions are found by diagonalizing Eq. 1 and only depend on the $\boldsymbol{n}_\pm \cdot \boldsymbol{\tau}$ term because the first term is proportional to the identity. Since $\boldsymbol{n}_\pm$ depends only on crystal angle $\theta$ and $\gamma_3/\gamma_2$, an experimental measurement of $\gamma_3/\gamma_2$ allows the reconstruction of the Bloch wavefunctions. Even for $k_z \neq 0$, although the Luttinger Hamiltonian is not block diagonal, knowing $\gamma_3/\gamma_2$ is still sufficient to reconstruct the Bloch wavefunctions.

**Reconstruction of Bloch Wavefunctions**

We use ratios of $T$-matrix elements to check the validity of the theory and measure $\gamma_3/\gamma_2$. Because the diagonal elements of Eq. 1 are real, when sideband and NIR laser polarizations are the same, for each pathway producing a $\sigma^+_{HSG}$ photon there is an equivalent pathway producing a $\sigma^-_{HSG}$ photon (Fig. 2, Supplementary Discussion, Extended Data Fig. 4) through states related by time-reversal symmetry. Therefore, the ratio of diagonal dynamical Jones matrix elements *for all sideband indices and crystal angles* is

$$\frac{T_{++,n}(\theta)}{T_{--,n}(\theta)} \equiv \xi_n(\theta) = 1 \quad \text{(Eq. 3)}$$

Because the off-diagonal elements of Eq. 1 are complex, when sideband and NIR laser polarizations are different, for each pathway producing a $\sigma^+_{HSG}$ photon there is an equivalent pathway producing a $\sigma^-_{HSG}$ photon with a complex-conjugated phase factor (Supplementary Discussion). Therefore, the ratio of off-diagonal dynamical Jones matrix elements *for all sideband indices* is

$$\frac{T_{+-,n}(\theta)}{T_{-+,n}(\theta)} \equiv \chi_n(\theta) = \frac{\gamma_2 \sin(2\theta) - i\gamma_3 \cos(2\theta)}{\gamma_2 \sin(2\theta) + i\gamma_3 \cos(2\theta)} \quad \text{(Eq. 4)}$$

The magnitude of $\chi_n(\theta)$ is 1 for all angles, but the argument depends on $\gamma_3/\gamma_2$ and $\theta$.

The experimentally measured values $\chi_n(\theta)$ and $\xi_n(\theta)$ at various $\theta$ are compared with the predictions of Eqs. 3 and 4 in Fig. 3 and Extended Data Fig. 5 using values for $\gamma_3$ and $\gamma_2$ recommended in[13]. Within experimental error, $|\chi_n(\theta)|$ and $|\xi_n(\theta)|$ are 1, as predicted by Eqs. 3 and 4 (Fig. 3a, Extended Data Fig. 5b,c). The arguments of $\chi_n(\theta)$ for eight different $\theta$ are independent of $n$ (Fig. 3b), lying within 22° of the constant values predicted by Eq. 4 (dashed lines) for all $\theta$ except $-45°$. The values of $\chi(\theta) \equiv \langle \chi_n(\theta) \rangle$ and $\xi(\theta) \equiv \langle \xi_n(\theta) \rangle$, where averages are over $n$, are plotted at each $\theta$ in Figs. 3c and d. The magnitudes $|\chi(\theta)|$ and $|\xi(\theta)|$ are independent of $\theta$, with a value of 1, as predicted by Eqs. 3 and 4 (Fig. 3c). The argument of $\chi(\theta)$ is plotted with respect to $\theta$ in Fig. 3d, and is close to the prediction provided by Eq. 4. Averaging the argument of $\chi(\theta)$ over experimentally sampled $\theta$ gives $\gamma_3/\gamma_2 = 1.47 \pm 0.48$, within experimental error of the value 1.42 recommended in[13]. We attribute the deviations in measured $\chi_n(\theta)$ and $\xi_n(\theta)$ from theoretical predictions, as well as much of the experimental error in the determination of $\gamma_3/\gamma_2$, to small inhomogeneous strain in the GaAs membrane (Methods, Extended Data Fig. 2).

From $\gamma_3/\gamma_2$, we reconstruct the Bloch wavefunctions of the Luttinger Hamiltonian in GaAs. For two coupled bands, the Bloch wavefunctions can be represented as spinors on a Bloch sphere[26]. In the $k_z = 0$ plane, each block of the Luttinger Hamiltonian is a two-by-two matrix, whose eigenfunctions—the Bloch wavefunctions—depend on $\theta$ but not on $|\mathbf{k}| = k$. Thus, in the $k_z = 0$ plane, for any $\theta$, a single point on the Bloch sphere represents the Bloch wavefunctions for arbitrary $k$. The closed black curves in the Northern and Southern hemispheres of the Bloch sphere in Fig. 4 represent the most likely Bloch wavefunctions consistent with our measured $\gamma_3/\gamma_2$ for the LH and HH, respectively. The North and South poles represent the states with spin -3/2 and +1/2, respectively. The Bloch wavefunctions for the degenerate partners of those represented in Fig. 4 are related by time-reversal symmetry.

**Discussion**

The complete electronic structure of a crystalline solid should include both its band structure and Bloch wavefunctions. We have reconstructed low-energy Bloch wavefunctions of holes in GaAs from polarimetry of high-order sideband spectra. GaAs is one of the most widely studied semiconductors, and the consistency of our results with the vast body of complementary previous work validates the novel method presented here. HSG can, in principle, be observed from any direct-gap semiconductor or insulator, and has been observed in semiconductor quantum wells[2,3,6,7] and both monolayer and bulk semiconducting transition metal dichalchogenides[5,8,9]. Thus, we expect polarimetry of high-order sidebands can be measured from a large class of bulk and nanostructured materials. As a probe of electronic structure, HSG's sensitivity to the bulk of electrically-insulating materials has the potential to complement ARPES, which works best on surfaces.

Holes in GaAs are an interesting special case because the Luttinger Hamiltonian is one of the simplest non-trivial low-energy effective Hamiltonians in solids. However, HSG spectra contain a wealth of information about the portions of the Brillouin zone explored during the acceleration phase, and straightforward extensions of the work presented here will enable reconstruction of Bloch wavefunctions from a wide range of low-energy Hamiltonians. Each peak in a HSG spectrum can be thought of as the output of an interferometer for Bloch waves. During acceleration by the THz field, Bloch wavepackets generally accumulate two types of phases: dynamic phases $A_{HH(LH)}$, which have been extensively discussed in this Article, and geometric phases (also called Berry phases)[6]. Dynamic phases depend only on the time-dependent energy eigenvalues of electrons and holes during acceleration. Geometric phases accumulate if the Bloch wavefunctions of electrons or holes change along their trajectories. The Luttinger model predicts that, along special trajectories which are straight lines through the Gamma point oriented along the constant angle $\theta$ defined by the linearly polarized THz electric field, electron-hole pairs acquire dynamic phases $A_{HH(LH)}$ but no Berry phases (Supplementary Discussion, Extended Data Fig. 7). Experimental conditions (linearly-polarized THz electric field and NIR excitation with $k = 0$) were chosen to excite only such trajectories in order to simplify analysis. However, in Hamiltonians with lower symmetry, even for these simple experimental conditions, Berry phases will, in general, accumulate during acceleration. For example, the addition of a small biaxial strain to GaAs lowers the symmetry, splits degeneracy between HH and LH bands at the Gamma point, and makes it inevitable that both dynamic and non-Abelian Berry phases contribute to the intensity and polarization of sidebands (Supplementary Discussion, Extended Data Fig. 8). We are currently investigating the hypothesis that the imperfect agreement between experiment and theory in Fig. 3 is due to small strains in this sample.

We expect Bloch wavefunctions of nearly-degenerate bands can be reconstructed for a wide range of semiconducting and insulating materials by minimizing the difference between measured Jones matrix elements and those predicted by the appropriate effective Hamiltonian. We anticipate the following requirements on materials and light sources to enable Bloch wavefuction reconstruction. The carrier density should be sufficiently low that it does not interfere with the recollision process. One must know the space group of the material and the spin and orbital angular momenta associated with each of the optically-excited bands at the band edge in order to construct a low-energy effective Hamiltonian with the correct symmetry and set of parameters using $\boldsymbol{k} \cdot \boldsymbol{p}$ theory. The laser tuned to the band gap of the material should be sufficiently weak that sideband intensities are in the linear regime. The strong THz laser should have sufficiently narrow bandwidth and low enough frequency to resolve multiple sidebands in the portion of the Brillouin zone of interest. The THz electric field should be strong enough to ionize electron-hole pairs in a fraction of a cycle, and for acceleration and recollision to occur for a detectable fraction of photoexcited quasiparticles before scattering disrupts the process, but not so strong that it creates electron-hole pairs spontaneously. By leveraging a variety of state-of-the-art narrow-band THz sources[27], we believe that Bloch wavefunction reconstruction from polarimetry of high-order sidebands can become an important new technique for determining the complete low-energy electronic structure of charged quasiparticles in both weakly and strongly-correlated materials.

**Acknowledgments** We gratefully acknowledge Profs. Renbao Liu and Mackillo Kira for reading an earlier version of the manuscript; Dr. Garrett Cole and Paula Heu for assistance with GaAs membrane fabrication; Mr. Alex Peñaloza for assistance with design and fabrication of cryostat modifications; Mr. Cameron Cannon for implementing software for Monte Carlo error estimation; Mr. David Enyeart and Dr. Nikolay Agladze for assistance with maintaining and operating the UCSB MM-wave free-electron laser; and Jerry Meyer and Igor Vurgaftman for a discussion. The portion of this research conducted at UCSB was funded by NSF-DMR 1710639 and NSF-DMR 2004995. Upgrades to the UCSB Terahertz facility that was used for this research were funded by NSF-DMR 1626681 and NSF-DMR 1126894. The portion of this research conducted at Princeton was funded in part by the Gordon and Betty Moore Foundation's EPiQS Initiative, Grant GBMF9615 to L. N. Pfeiffer, and by the National Science Foundation MRSEC grant DMR 1420541.

**Author contributions** Performing experiments, data collection, and analysis: S. O. and J. C. Software: J. C., D. V. and S. O. Theory: Q. W. Conceptualization: M. S. S. and Q. W. Resources (GaAs epilayer growth): K. W. W. and L. P. Development of broad-band polarimetry: D. V.

Writing: M. S. S., Q. W., S. O. and J. C. Supervision, funding acquisition and project administration: M. S. S.

**Competing interests:** The authors declare no competing interests.

**Additional information**

**Supplementary information** is available for this paper.
**Correspondence and requests for materials** should be addressed to M. S. S.

# Main Figures

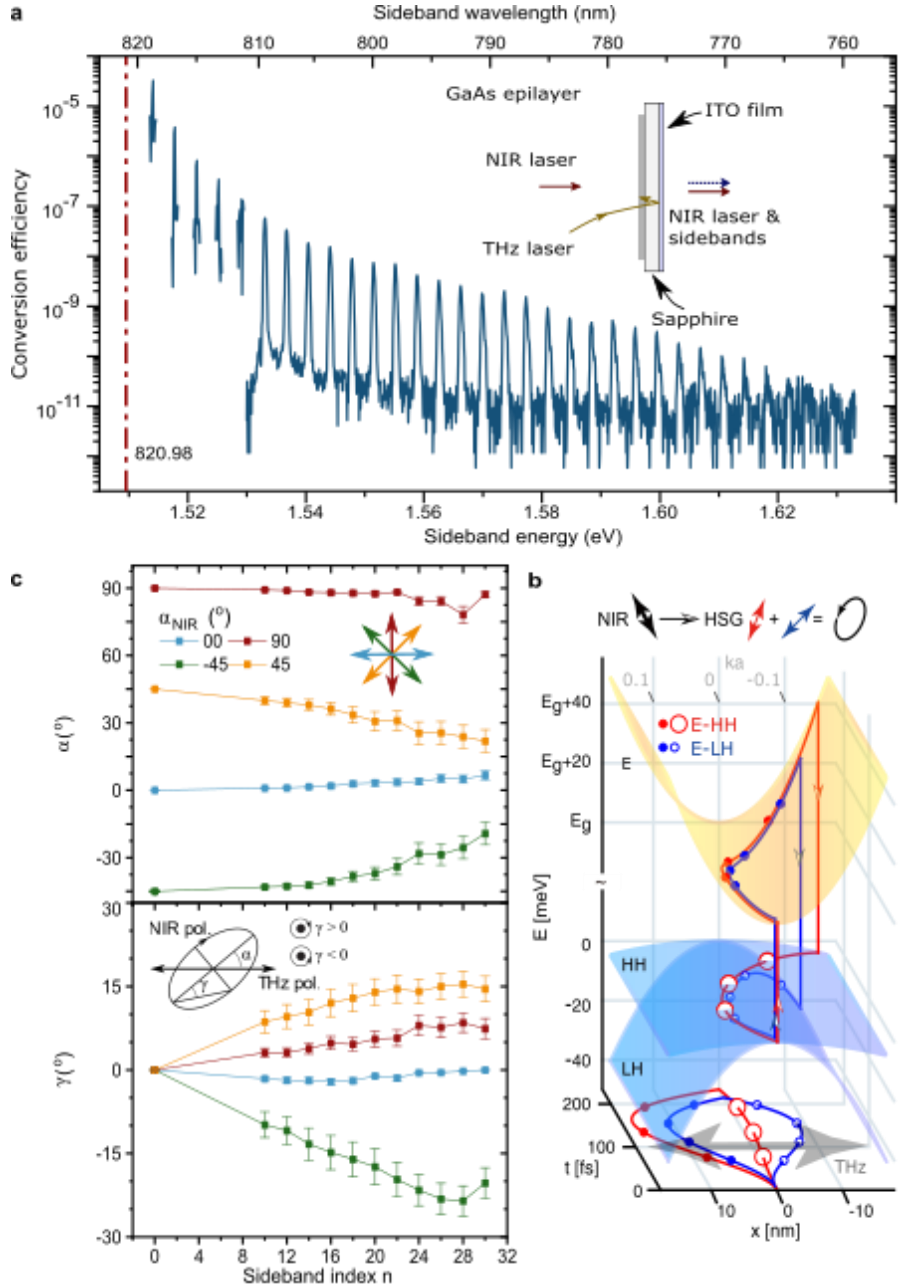

**Fig. 1 | High-order sideband generation (HSG) in bulk GaAs. a**, An HSG spectrum (blue). The red dash-dot line indicates the NIR laser photon energy. Sidebands (peaks) are spaced by twice the THz frequency. **Inset**, Experimental setup schematic. **b,** Three-step model of HSG in GaAs. Electron-heavy hole (E-HH) and electron-light hole (E-LH) recollision trajectories are denoted in red and blue, respectively, and are classical representations of interfering quantum pathways. The y-axis (into page) corresponds to time. Bottom plane: real-space trajectories of E-HH and E-LH pairs. 3-D mesh plot: $k$-space trajectories of the same pairs. The $z$-axis corresponds to energy, and the $x$-axis corresponds to dimensionless quasimomentum $ka$, where $a$ is the lattice constant. Step 1: creation of E-HH and E-LH pairs by NIR laser (up arrows of equal length). Step 2:

acceleration by the THz field. Electrons and holes begin at $k = 0$, and recollide with substantial kinetic energy at $|ka| > 0.1$. Step 3: recombination of E-HH and E-LH pairs and emission of sideband (down arrows of equal length). Top line: linearly-polarized NIR laser photons (black double arrow) lead to emission of E-HH and E-LH sideband components (red and blue double arrows) with rotated linear polarizations and different phases, which combine to emit an elliptically-polarized sideband. Classical calculations of trajectories (Methods) are for the 24$^{th}$ order sideband. **c,** Sideband linear orientation angle $\alpha$ and ellipticity angle $\gamma$ as functions of sideband index for NIR laser linear orientation $\alpha_{NIR} = 0°$ (cyan), 90° (red), 45° (orange), and $-45°$ (green) defined in upper inset. NIR laser polarization angles are plotted as sidebands with $n = 0$. The lower inset defines $\alpha$ and $\gamma$ with respect to the linearly-polarized THz field. The measured polarization at each sideband index is displayed directly below the corresponding peak in the HSG spectrum **a**. The error bars denote the standard deviation.

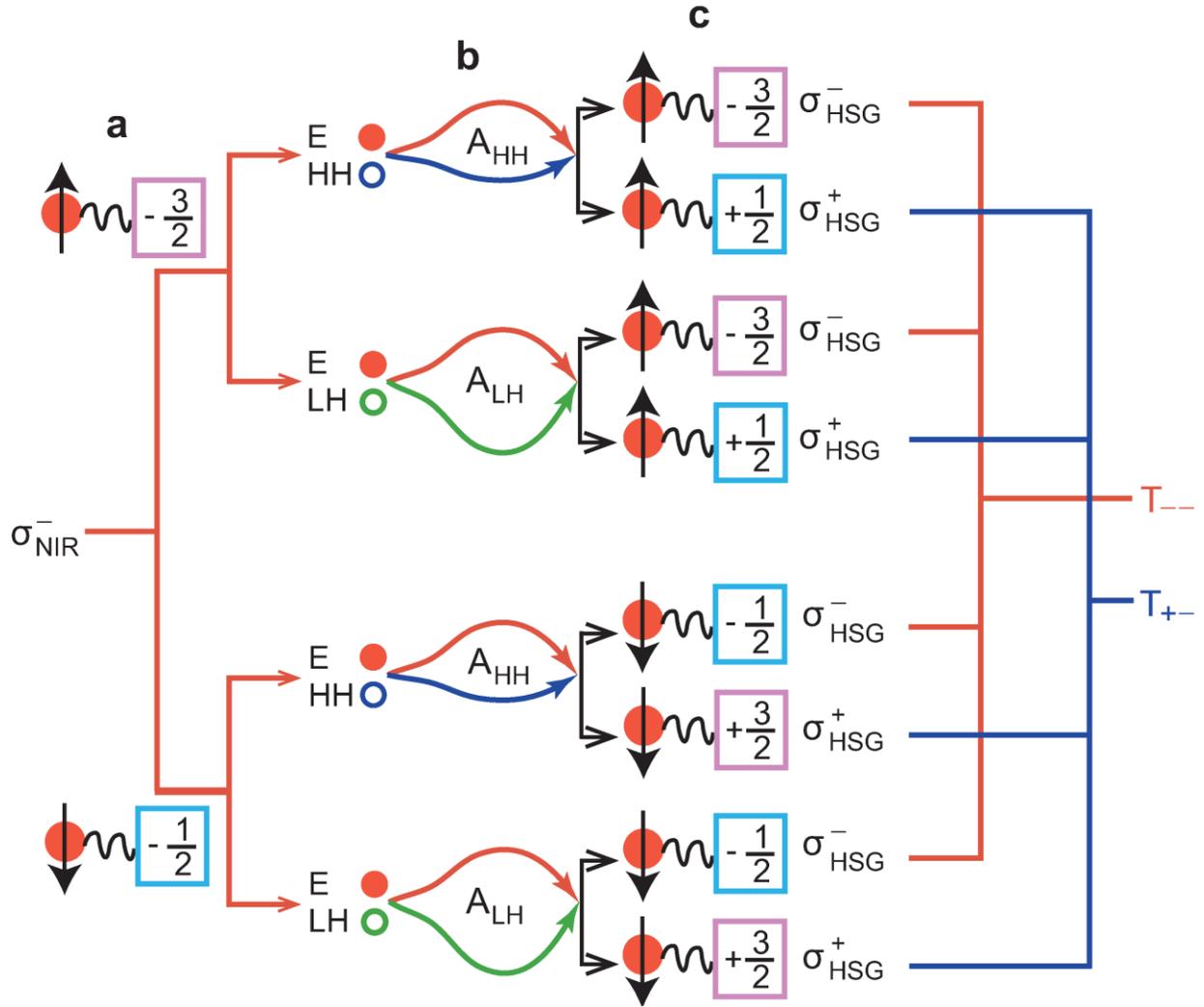

**Fig. 2 | Quantum interference leading to sideband polarization.** A photon from the NIR laser is decomposed into components $\sigma_{NIR}^{\pm}$, with helicity $\pm 1$. **a,** A $\sigma_{NIR}^{-}$ photon excites either a spin-up electron and hole of spin -3/2 or a spin-down electron and hole of spin -1/2. **b,** Driven by the THz field, an electron-hole pair accumulates dynamic phase $A_{HH}$ or $A_{LH}$, depending on the band of the hole state (HH or LH). The electron spin is unchanged, while the hole states originating from the spin -3/2 state are superpositions of spin -3/2 and +1/2 states and the states originating from the spin -1/2 state are superpositions of spin -1/2 and +3/2 states. **c,** Upon recollision, either $\sigma_{HSG}^{+}$ or $\sigma_{HSG}^{-}$ photons are produced following angular momentum conservation—for example, a spin +3/2 hole recombining with a spin-down (-1/2) electron produces a $\sigma_{HSG}^{+}$ photon with helicity +3/2 -1/2 = +1. The interference of the evolution pathways from $\sigma_{NIR}^{-}$ to $\sigma_{HSG}^{+}$ ($\sigma_{HSG}^{-}$) produces the dynamical Jones matrix element $T_{+-}$ ($T_{--}$). Photons with $\sigma_{NIR}^{+}$ result in similar pathways to produce to $T_{-+}$ and $T_{++}$ (Supplementary Discussion, Extended Data Fig. 4).

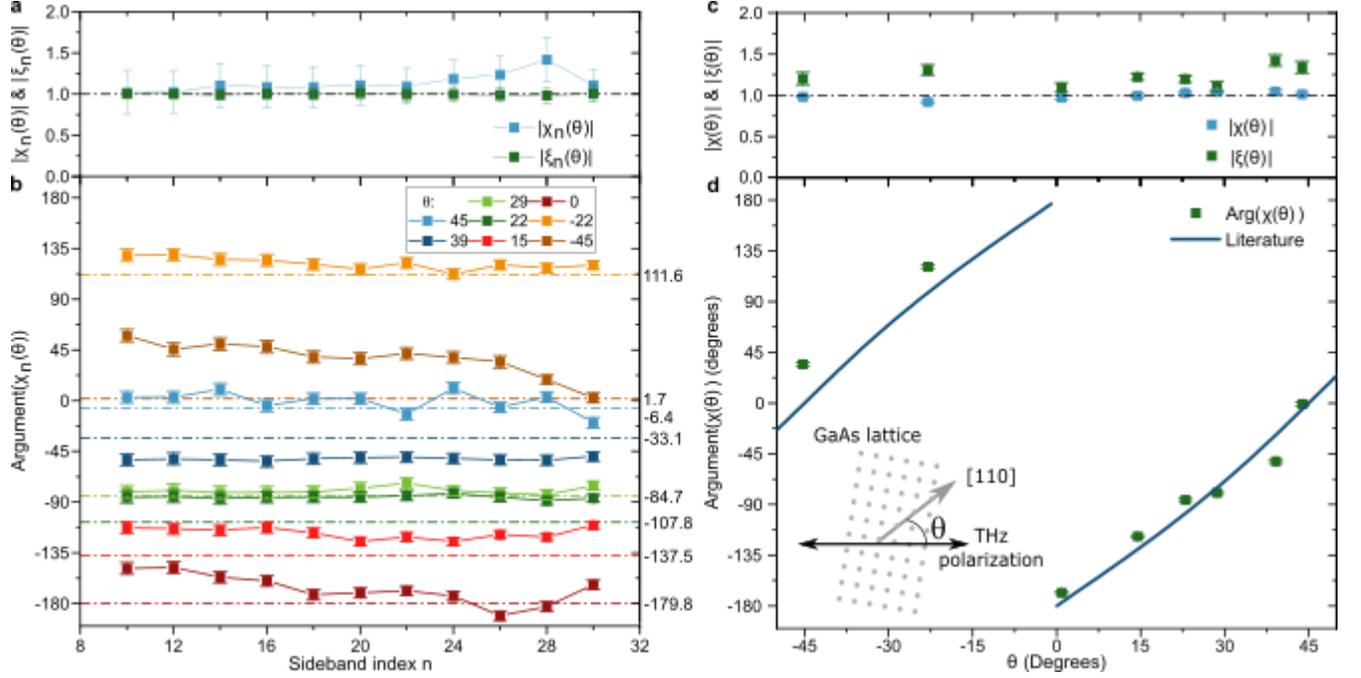

**Fig. 3 | Ratios of Jones matrix elements, $\xi_n(\theta) \equiv T_{++,n}(\theta)/T_{--,n}(\theta)$ and $\chi_n(\theta) \equiv T_{+-,n}(\theta)/T_{-+,n}(\theta)$, measured by Stokes polarimetry. a**, The magnitude of $\chi_n(\theta)$ and $\xi_n(\theta)$ at $\theta = 39°$. The dash-dot line marks the magnitude of 1 predicted by Eq. 3. For other $\theta$, see Extended Data Fig. 5. **b,** The argument of $\chi_n(\theta)$ as a function of sideband index $n$, at various $\theta$. The dash-dot lines mark the expected values (noted on the right) from Eq. 4, using values of $\gamma_2$ and $\gamma_3$ recommended in [13]. **c**, The magnitudes of $\chi(\theta) \equiv \langle\chi_n(\theta)\rangle$ and $\xi(\theta) \equiv \langle\xi_n(\theta)\rangle$ averaged over sideband index $n$. The dash-dot line indicates the magnitude of 1 predicted by Eqs. 3 and 4. **d**. The argument of $\chi(\theta)$ at each experimentally probed $\theta$. The solid blue line is the expected argument from Eq. 4 using the values of $\gamma_2$ and $\gamma_3$ recommended in [13]. **Inset**, The definition of $\theta$, using the GaAs crystal lattice and the THz electric field. Error bars denote one standard deviation.

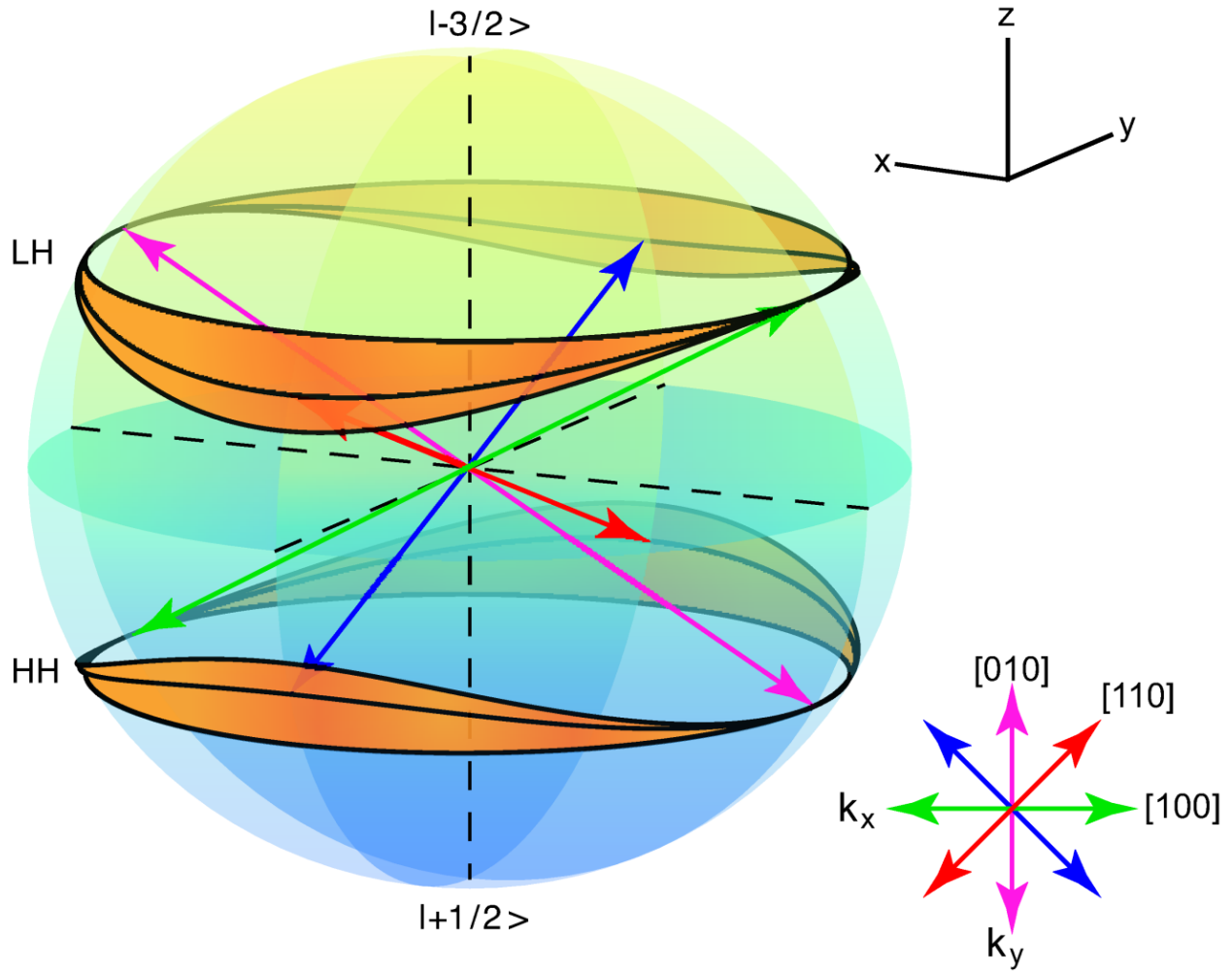

**Fig. 4 | Reconstruction of the Bloch wavefunctions for $k_z = 0$.** The Bloch wavefunctions of heavy hole (HH) and light hole (LH) bands associated with $H_+$ in Eq (1) are plotted as black lines. The orange shaded area corresponds to the uncertainty in the wavefunction associated with one standard deviation in the measurement of $\gamma_3/\gamma_2$. For a given $\theta$, each wavefunction is represented by a point on the Bloch sphere. The arrows within the Bloch sphere point from the origin to the LH and HH Bloch wavefunctions for the values of $\theta$ defined by their Miller indices in the inset below. The poles correspond to the spin -3/2 and spin +1/2 states. The wavefunctions for $H_-$ are paths reflected across the $xz$-plane on a Bloch sphere with poles representing the spin +3/2 and spin -1/2 states.

# Methods

**Fabrication of GaAs Sample**

A 500 nm thick GaAs epilayer was grown via molecular beam epitaxy (MBE) and then transferred onto a sapphire substrate through Van-der-Waals bonding[28-31]. The sapphire was transparent to both near-infrared (NIR) and terahertz (THz) radiation. The fact that the thermal expansion coefficients of sapphire and GaAs are closely matched ensures relatively small strains in the GaAs epilayer upon thermal cycling. In order to make the strain as small and homogeneous as possible, the GaAs epilayer was etched to be circular. A layer of indium tin oxide (ITO)—which reflects THz while transmitting NIR radiation—was grown on the sapphire surface that was opposite to the GaAs epilayer to create a low-Q cavity that enhanced the THz field in GaAs at selected THz frequencies. At the 447 GHz frequency used in this Article, the THz field is enhanced by a factor of 1.5 from the ITO layer (Extended Data Fig. 1). A silicon dioxide ($SiO_2$) anti-reflection coating was grown on top of the ITO to minimize its NIR reflection and avoid NIR Fabry-Perot oscillations in the sideband spectra. See Supplementary Methods for a step-by-step fabrication procedure.

Absorbance spectra of the GaAs epilayer were measured in a cryogenic chamber as a preliminary characterization on strains, as well as the excitation gap, which motivated our choice of NIR laser wavelengths for HSG experiments. Extended Data Fig. 2 shows an absorbance spectrum measured at a sample temperature of 60 K using a white light source, and calculated as A=-10 Log (Transmitted power with sample in cryostat/Transmitted power with cryostat (and sample) removed). The sharp peaks are assigned to exciton resonances associated with band-edge states with different angular momenta. These peaks are separated by 2.6 meV. A recent study has associated a similar splitting with a strain of order 0.1%[32]. The absorbance spectra in the immediate neighborhood of the illuminated spot chosen for the HSG experiments in this Article showed little variation.

**Optical methods**

The NIR laser was generated from an M Squared SolTiS® Ti:Sapphire laser, with a 7 W, 532 nm Sprout laser as the pump. The M Squared cavity is tunable via piezoelectric response, with a precision of 0.01 Å output NIR wavelength, measured in real time by a WS6-600 wavemeter. The linewidth of the SolTiS is <5 MHz, enabling excitation of electron-hole pairs with very well-defined energy and contributing negligibly to sideband linewidth, which is determined primarily by pulse-to-pulse fluctuations in the FEL frequency[7]. An acousto-optic modulator was used to direct the NIR laser onto the sample for 1 µs at a 0.0001% duty cycle, synchronized with the THz output pulse from the FEL. After the modulation, only the 1st order beam propagated through the rest of the optical elements. The polarization of the NIR laser beam incident onto the GaAs epilayer was set with a quarter-wave plate and a half-wave plate, and measured by a Thorlabs PAX polarimeter. The NIR beam was focused down to ~500 µm at the GaAs epilayer using a 500 mm lens.

The THz radiation was generated from the cavity-dumped UCSB MM-wave free-electron laser[24,33,34]. Most of the variance in the output frequency was due to variance in the terminal voltage of the electrostatic accelerator that drives the FEL[7]. The THz beam output from an optical transport system was split into two beam paths. 10% of the THz output power was directed into a fast-response pyroelectric reference detector, which measured the output power of each FEL pulse. The other 90% of the THz output power was directed onto the cryostat containing the GaAs epilayer. A 12.5 cm, gold-coated off-axis parabolic mirror was used to focus the THz beam into a 1.2 mm diameter spot. An ITO slide, which was transmissive in NIR but reflective in THz range, was used to adjust the THz beam spot on the GaAs epilayer and make sure the NIR and THz fields were collinear. The pulse energy was measured on each day prior to the HSG experiments using a Thomas Keating absolute power/energy meter placed after the beam splitter but before the parabolic mirror. The pulse energy measured by the Thomas Keating power meter was used to calibrate the fast pyroelectric reference detector.

The THz field strength in the GaAs epilayer is estimated to be 70±2 kV/cm. This high THz field strength resulted in the large number of sidebands reported in this Article. However, the dependence of sideband polarization angles on NIR laser polarization angles measured at fields as low as 35 kV/cm was similar to the dependence reported in Fig. 1c. In the calculation of THz field strength, we assume that the gold-coated off axis parabolic mirror is 100% reflective, the ITO slide is 70% reflective, the cryostat window is 95% transmissive, and the ITO coating on the sapphire provides a 150% field enhancement on the sample.

The sidebands generated from the GaAs sample were first transmitted through a Stokes polarimeter, which includes a rotating quarter-wave plate (RQWP) and a horizontal linear polarizer. The Stokes polarimeter was calibrated by measuring the NIR laser polarizations with the Thorlabs PAX polarimeter, which was impractical for Stokes polarimetry of the sidebands because it is optimized for use with a cw laser beam at a single frequency. The intensity of each sideband was measured either by a photomultiplier tube (PMT) or a charge-coupled device (CCD), each coupled to a dedicated monochromator[6]. The PMT measured the lowest-order sidebands, while the CCD imaged many higher-order sidebands simultaneously. To optimize the efficiencies of the diffraction gratings, a half-wave plate was placed after the Stokes polarimeter to rotate the sideband polarizations.

All measurements were performed at 60 K, which was the base temperature of the cryostat during this experimental campaign. HSG polarimetry spectra recorded at lower temperatures were similar to those reported here.

**Extraction of $\gamma_3/\gamma_2$ from Stokes polarimetry**

We characterize the polarization of each sideband using the four Stokes parameters defined as $S_0 = I$, $S_1 = Ip\cos 2\alpha \cos 2\gamma$, $S_2 = Ip\sin 2\alpha \cos 2\gamma$, and $S_3 = Ip\sin 2\gamma$, where $I$ is the total intensity, $p$ is the degree of polarization, and the orientation angle $\alpha$ and ellipticity angle $\gamma$ are defined in the inset of Extended Data Fig. 3b. After a sideband passes through the RQWP and horizontal linear polarizer, the intensity of the outgoing light, $S_{out}(\phi)$, can be expressed as

$$S_{out}(\phi) = \frac{S_0}{2} + \frac{S_1}{4} - \frac{S_3}{2}\sin(2\phi) + \frac{S_1}{4}\cos(4\phi) + \frac{S_2}{4}\sin(4\phi)$$

where $\phi$ is the angle between the fast-axis of the RQWP and the horizontal. By measuring $S_{out}$ as a function of $\phi$, the four Stokes parameters can be extracted from the Fourier transform $\mathcal{F}_m = \int_0^{2\pi} S_{out}(\phi)e^{-im\phi}d\phi/2\pi$ : $S_0 = 2\mathcal{F}_0 - 4\text{Re}(\mathcal{F}_4)$, $S_1 = 8\text{Re}(\mathcal{F}_4)$, $S_2 = -8\text{Im}(\mathcal{F}_4)$, $S_3 = 4\text{Im}(\mathcal{F}_2)$. We sampled the intensities of each sideband at 16 different angles $\phi$. We define plots of $S_{out}$ as functions of the angle $\phi$ as "polaragrams" (see Extended Data Fig. 3a and c for examples). For each angle $\phi$, four CCD scans were taken to establish the variance of the intensity $S_{out}$. From the Stokes parameters of the $n$-th order sideband, $S_{i,n}$, the polarization state of the sideband can be extracted by calculating the angles $\alpha_n$ and $\gamma_n$ from relations

$$\tan(2\alpha_n) = \frac{S_{2,n}}{S_{1,n}}$$

$$\tan(2\gamma_n) = \frac{S_{3,n}}{\sqrt{S_{1,n}^2 + S_{2,n}^2}}$$

Examples of extracted polarization states of sidebands are shown in Extended Data Fig. 3b and d. To reconstruct the dynamical Jones matrices, the polarization states of the sidebands were measured for four different polarization states of the NIR laser. All polarizations of the NIR laser were linear ($\gamma_{\text{NIR}} = 0°$) with orientation angles $\alpha_{\text{NIR}} = 0°, 45°, 90°$, and $-45°$, respectively. Each dynamical Jones matrix $\mathcal{J}$ connects the electric fields of the NIR laser and a sideband through

$$\begin{pmatrix} E_{x,n} \\ E_{y,n} \end{pmatrix} = \begin{pmatrix} \mathcal{J}_{xx,n} & \mathcal{J}_{xy,n} \\ \mathcal{J}_{yx,n} & \mathcal{J}_{yy,n} \end{pmatrix} \begin{pmatrix} E_{x,\text{NIR}} \\ E_{y,\text{NIR}} \end{pmatrix}$$

where

$$\begin{pmatrix} E_{x,n} \\ E_{y,n} \end{pmatrix} = \begin{pmatrix} \cos \alpha_n & -\sin \alpha_n \\ \sin \alpha_n & \cos \alpha_n \end{pmatrix} \begin{pmatrix} \cos \gamma_n \\ i \sin \gamma_n \end{pmatrix} \equiv \begin{pmatrix} \cos \beta_n \\ e^{i\delta_n} \sin \beta_n \end{pmatrix} e^{i\zeta_n}$$

$$\begin{pmatrix} E_{x,\text{NIR}} \\ E_{y,\text{NIR}} \end{pmatrix} = \begin{pmatrix} \cos \alpha_{\text{NIR}} & -\sin \alpha_{\text{NIR}} \\ \sin \alpha_{\text{NIR}} & \cos \alpha_{\text{NIR}} \end{pmatrix} \begin{pmatrix} \cos \gamma_{\text{NIR}} \\ i \sin \gamma_{\text{NIR}} \end{pmatrix} \equiv \begin{pmatrix} \cos \beta_{\text{NIR}} \\ e^{i\delta_{\text{NIR}}} \sin \beta_{\text{NIR}} \end{pmatrix} e^{i\zeta_{\text{NIR}}}$$

The ratio $E_{y,n}/E_{x,n}$ yields the equation

$$\cos \beta_n \left( \frac{\mathcal{J}_{yx,n}}{\mathcal{J}_{xx,n}} \cos \beta_{\text{NIR}} + \frac{\mathcal{J}_{yy,n}}{\mathcal{J}_{xx,n}} e^{i\delta_{\text{NIR}}} \sin \beta_{\text{NIR}} \right) - e^{i\delta_n} \sin \beta_n \left( \cos \beta_{\text{NIR}} + \frac{\mathcal{J}_{xy,n}}{\mathcal{J}_{xx,n}} e^{i\delta_{\text{NIR}}} \sin \beta_{\text{NIR}} \right) = 0$$

which is linear with respect to the ratios $\mathcal{J}_{yx,n}/\mathcal{J}_{xx,n}$, $\mathcal{J}_{yy,n}/\mathcal{J}_{xx,n}$, and $\mathcal{J}_{xy,n}/\mathcal{J}_{xx,n}$. Measurements for three polarization states of the NIR laser give three such linear equations, which uniquely determine the ratios between the dynamical Jones matrix elements. From the measurements for the four NIR polarizations, we obtained four linear equations, which were solved by the method of least squares. The absolute values of the dynamical Jones matrix elements, which are not concerns of this Article, can be determined through the absolute values of the Stokes parameters.

Each dynamical Jones matrix $\mathcal{J}$ was converted to the $T$-matrix in a basis of circular polarizations through unitary transformation $T = U^\dagger \mathcal{J} U$, where

$$U = \frac{1}{\sqrt{2}} \begin{pmatrix} e^{-i\varphi} & -e^{i\varphi} \\ ie^{-i\varphi} & ie^{i\varphi} \end{pmatrix}$$

Here, $\varphi$ is the angle between the THz polarization and the [100] crystal direction.

From Eq. 4 in the main text, with the measured $T$-matrix, the ratio $\gamma_3/\gamma_2$ can be calculated as

$$\frac{\gamma_3}{\gamma_2} = |\tan 2\theta| \sqrt{\frac{1 - \cos Arg(T_{+-,n}/T_{-+,n})}{1 + \cos Arg(T_{+-,n}/T_{-+,n})}}$$

where $Arg(T_{+-,n}/T_{-+,n})$ is the argument of $T_{+-,n}/T_{-+,n}$, and $\theta$ is the angle between the THz polarization and the [110] crystal direction. From each angle $\theta$ ($\sin 2\theta, \cos 2\theta \neq 0$) and the ratio $T_{+-,n}/T_{-+,n}$ for each sideband, one value of $\gamma_3/\gamma_2$ was obtained. An average over sideband index $n$ and angle $\theta$ yields $\gamma_3/\gamma_2 = 1.47 \pm 0.48$, where the quoted error is the standard deviation of $\gamma_3/\gamma_2$.

A Monte Carlo simulation was performed to estimate the errors in the dynamical Jones matrix elements from two sources which were added in quadrature: (1) the variance in the sideband intensity measurements and (2) the deviation $\delta\eta$ of RQWP retardance from its ideal value $\pi/2$. A small deviation $\delta\eta$ in the RQWP retardance modifies the relations between the Fourier transforms $\mathcal{F}_m$ and Stokes parameters as: $S_0 = 2\mathcal{F}_0 - 4\text{Re}(\mathcal{F}_4)(1-\delta\eta)/(1+\delta\eta)$, $S_1 = 8\text{Re}(\mathcal{F}_4)/(1+\delta\eta)$, $S_2 = -8\text{Im}(\mathcal{F}_4)/(1+\delta\eta)$, $S_3 = 4\text{Im}(\mathcal{F}_2)/(1-\delta\eta^2/2)$. The deviation $\delta\eta$ was calibrated to be in the range $[-\pi/36, \pi/36]$. The angles $\alpha_n$ and $\gamma_n$ of the $n$-th order sideband were randomly sampled from normal distributions, with the mean and standard deviation set as the measured mean values and errors as shown in Extended Data Fig. 6a. Each set of $\alpha_n$ and $\gamma_n$ were sampled 10,000 times, generating 10,000 values for each of the dynamical Jones matrix elements. As an example, Extended Data Fig. 6b shows the distribution of 1000 sets of $\alpha_{12}$ and $\gamma_{12}$ for four polarization states of the NIR laser. The value and error of each dynamical Jones matrix element was calculated as the mean and the standard deviation of the generated distribution, respectively. Note that the dynamical Jones matrix elements are complex valued, and we set $\mathcal{J}_{xx,n} = 1$ in this Article. Extended Data Fig. 6c shows the distributions of the dynamical Jones matrix elements produced from the distributions of $\alpha_{12}$ and $\gamma_{12}$ in Extended Data Fig. 6b.

**Effective Hamiltonian and low-energy Bloch wavefunctions**

In the basis consisting the eigenvectors of the position operator $\hat{r}$, a Bloch wavefunction with band index $N$ and quasimomentum $\boldsymbol{k}$ has the form $\langle r|\psi_{N,k}\rangle = e^{i\boldsymbol{k}\cdot\boldsymbol{r}}u_{N,k}(\boldsymbol{r})$, where $u_{N,k}(\boldsymbol{r}) = \langle r|u_{N,k}\rangle$ is a periodic function with the periodicity of the crystal. The eigenvalue problem of the Bloch wave functions, $H|\psi_{N,k}\rangle = E_{N,k}|\psi_{N,k}\rangle$, can be equivalently stated as an eigenvalue problem of the state $|u_{N,k}\rangle$ in the form $H(\boldsymbol{k})|u_{N,k}\rangle = E_{N,k}|u_{N,k}\rangle$, where $H(\boldsymbol{k}) \equiv e^{-i\boldsymbol{k}\cdot\hat{r}}He^{i\boldsymbol{k}\cdot\hat{r}}$. According to the $\boldsymbol{k}\cdot\boldsymbol{p}$ method[35], in cases where the excited Bloch waves are located in energy bands that are relatively isolated and their quasimomenta are restricted in a small portion of the Brillouin zone, a finite number of states $|u_{N,k_0}\rangle$ at quasimomentum $\boldsymbol{k}_0$ can be approximately taken as a complete basis. On this finite basis, the Hamiltonian $H(\boldsymbol{k})$ can be represented as a finite-dimensional matrix—the effective Hamiltonian, whose eigenfunctions, the low-energy Bloch wavefunctions, are linear combinations of the states $\{|u_{N,k_0}\rangle\}$. Determination of the effective Hamiltonian does not rely on the exact representations of the states $\{|u_{N,k_0}\rangle\}$ in real-space coordinate but their symmetry properties. The Luttinger Hamiltonian is an effective Hamiltonian with the basis chosen as four valence-band-edge states (Supplementary Discussion).

**Interference of Bloch waves**

We consider the case where the photon energy of the NIR laser lies just below the bandgap and assume that the sideband amplitudes are dominantly determined by electron-hole pairs created at $\boldsymbol{k} = 0$. Under an approximation of free electrons and holes, the amplitude of the $n$-th sideband can be written as (Supplementary Discussion)

$$\mathbb{P}_n = \sum_{s=\pm} \frac{i\omega}{2\pi V\hbar} \int_0^{2\pi/\omega} dt\, e^{i(\Omega+n\omega)t} \int_{-\infty}^t dt'\, \mathbb{D}_s^\dagger R_s \begin{pmatrix} e^{iA_{HH}(t',t)} & 0 \\ 0 & e^{iA_{LH}(t',t)} \end{pmatrix} R_s^\dagger \mathbb{D}_s \cdot \boldsymbol{E}_{\text{NIR}}(t')$$

where $\omega$ is the angular frequency of the THz field, $V$ is the volume of the material, $\boldsymbol{E}_{\text{NIR}}(t) = \boldsymbol{F}_{\text{NIR}}e^{-i\Omega t}$ is the electric field of the NIR laser under the rotating wave approximation, the two components of $\mathbb{D}_+ = -d(\boldsymbol{\sigma}_-, \boldsymbol{\sigma}_+/\sqrt{3})^T$ ($\mathbb{D}_- = -d(\boldsymbol{\sigma}_+, \boldsymbol{\sigma}_-/\sqrt{3})^T$) are dipole matrix elements between spin-down (spin-up) electron and hole states with spin +3/2 (-3/2) and spin -1/2 (+1/2) respectively ($d$ is a constant dipole matrix element), and $R_\pm$ is a two-by-two unitary matrix that diagonalizes the hole Hamiltonian $[H_v^\pm(\boldsymbol{k})]^*$ through $R_\pm^\dagger(\hat{\boldsymbol{n}}_\pm \cdot \boldsymbol{\tau}^*)R_\pm = \tau_z$ with $\hat{\boldsymbol{n}}_\pm \equiv \boldsymbol{n}_\pm/|\boldsymbol{n}_\pm|$. The first and second column of $R_+$ ($R_-$) respectively represent the wavefunction of heavy-hole and light-hole on the basis of hole states with spin +3/2 (-3/2) and -1/2 (+1/2). The first (second) component of the quantity $R_\pm^\dagger \mathbb{D}_\pm \equiv (\mathfrak{D}_{HH,\pm}, \mathfrak{D}_{LH,\pm})^T$ represents the dipole matrix elements between E and HH (LH) bands. The acceleration process is described by the dynamic phase $A_{HH(LH)}(t',t)$, which contains the quasimomentum $\boldsymbol{k}_{t'}(t'') = e\boldsymbol{F}_{THz}(\sin\omega t' - \sin\omega t'')$ satisfying the initial condition $\boldsymbol{k}_{t'}(t') = \boldsymbol{0}$ indicated by the subscript $t'$ and $\hbar\partial_{t''}\boldsymbol{k}_{t'}(t'') = -e\boldsymbol{E}_{THz}(t'')$, with $e$ being the elementary charge and $\boldsymbol{E}_{THz}(t) = \boldsymbol{F}_{THz}\cos\omega t$ the THz electric field. The three-step process of HSG can thus be described as interference of the following recollision pathways: a Bloch wave associated with an electron-hole pair E-HH (E-LH) is first created by the NIR laser with amplitude proportional to $\mathfrak{D}_{HH(LH),\pm} \cdot \boldsymbol{E}_{NIR}(t')$, acquires a dynamic phase $A_{HH(LH)}(t',t)$ during the acceleration phase from $t'$ to $t$, and generates sidebands through the dipole vector $\mathfrak{D}_{HH(LH),\pm}$. The major contribution to the sideband amplitudes comes from the recollision pathways around the saddle-points $(t',t)$ given by the stationary-phase conditions:

$$-\hbar\frac{\partial A_{HH(LH)}(t',t)}{\partial t'} + \hbar\Omega = \int_{t'}^{t} dt'' \frac{\partial \boldsymbol{k}_{t'}(t'')}{\partial t'} \cdot (\nabla_k E_c[\boldsymbol{k}_{t'}(t'')] - \nabla_k E_{HH(LH)}[\boldsymbol{k}_{t'}(t'')]) = 0$$

$$-\hbar\frac{\partial A_{HH(LH)}(t',t)}{\partial t} = E_c[\boldsymbol{k}_{t'}(t)] - E_{HH(LH)}[\boldsymbol{k}_{t'}(t)] = \hbar\Omega + n\hbar\omega$$

We have used the condition $E_c(\boldsymbol{0}) - E_{HH(LH)}(\boldsymbol{0}) = E_g = \hbar\Omega$, where $E_g$ is the bandgap. Substituting the energy dispersion relations $E_c(\boldsymbol{k}) = E_g + \frac{\hbar^2 k^2}{2m_c}$ ($m_c$ is the effective mass of the conduction band), $E_{HH}(\boldsymbol{k}) = -\frac{\hbar^2 k^2}{2m_{HH}}$ and $E_{LH}(\boldsymbol{k}) = -\frac{\hbar^2 k^2}{2m_{LH}}$ ($m_{HH} = m_0(\gamma_1 - 2\gamma_2|\boldsymbol{n}_\pm|)^{-1}$ and $m_{LH} = m_0(\gamma_1 + 2\gamma_2|\boldsymbol{n}_\pm|)^{-1}$ are respectively the effective masses of the HH and LH bands) into the stationary-phase conditions, we obtain

$$\int_{t'}^{t} dt'' \boldsymbol{k}_{t'}(t'') = \boldsymbol{0}$$

$$\frac{\hbar^2 \boldsymbol{k}_{t'}^2(t)}{2\mu_{ch(cl)}} = n\hbar\omega$$

where $\mu_{ch} = \left(m_c^{-1} + \frac{\gamma_1 - 2\gamma_2|\boldsymbol{n}_\pm|}{m_0}\right)^{-1}$ ($\mu_{cl} = \left(m_c^{-1} + \frac{\gamma_1 + 2\gamma_2|\boldsymbol{n}_\pm|}{m_0}\right)^{-1}$) is the reduced mass of the E-HH (E-LH) pair. The first equation has the meaning that the electron-hole pairs return to the position where they are created. The second equation states energy conservation at recollision. For each sideband order $n$, these two equations can be solved for the saddle-points $(t',t)$, which determine k-space trajectories $\boldsymbol{k}_{t'}(t'')$, as well as classical real-space trajectories with the velocities of E, HH, and LH given by $\hbar\boldsymbol{k}_{t'}(t'')/m_c$, $-\hbar\boldsymbol{k}_{t'}(t'')/m_{HH}$, and $-\hbar\boldsymbol{k}_{t'}(t'')/m_{LH}$. Fig. 1b shows the shortest trajectory for the 24$^{th}$ order sideband, and parameters $m_c = 0.067m_0$, $m_{HH} = 0.711m_0$, and $m_{LH} = 0.081m_0$ are used in the calculation.

**Representation of Bloch wavefunctions**

The wavefunctions of the Hamiltonian $H_v^\pm(\boldsymbol{k})$ are eigenfunctions of $\hat{\boldsymbol{n}}_\pm \cdot \boldsymbol{\tau}$, which is defined on the basis of spin $\mp 3/2$ and $\pm 1/2$ states. We define $\hat{\boldsymbol{n}}_\pm = (\sin\Theta\cos\Phi, \sin\Theta\sin\Phi, \cos\Theta)$ as a point on a Bloch sphere with polar angle $\Theta$ and azimuthal angle $\Phi$, and write the eigenfunctions of $H_v^\pm(\boldsymbol{k})$ as:

$$|HH_\pm\rangle = \left(\cos\left(\frac{\Theta}{2}\right) \quad e^{i\Phi}\sin\left(\frac{\Theta}{2}\right)\right)$$

$$|LH_\pm\rangle = \left(-\sin\left(\frac{\Theta}{2}\right) \quad e^{i\Phi}\cos\left(\frac{\Theta}{2}\right)\right)$$

The point on the Bloch sphere with coordinates $\hat{\boldsymbol{n}}_\pm$ ($-\hat{\boldsymbol{n}}_\pm$) represents the state $|HH_\pm\rangle$ ($|LH_\pm\rangle$) for the HH (LH) band. The angles $\Theta$ and $\Phi$ are determined from the measured $\gamma_3/\gamma_2$ and angle $\theta$ through the definition $\boldsymbol{n}_\pm = \left(\frac{\sqrt{3}}{2}\sin(2\theta), \mp\frac{\sqrt{3}\gamma_3}{2\gamma_2}\cos(2\theta), -\frac{1}{2}\right)$.

**Data availability:** The datasets generated and/or analyzed during the current study are available in the xxx repository [36].

**Code availability:** The codes used in the data analysis are available at [37].

# Extended Data Figures

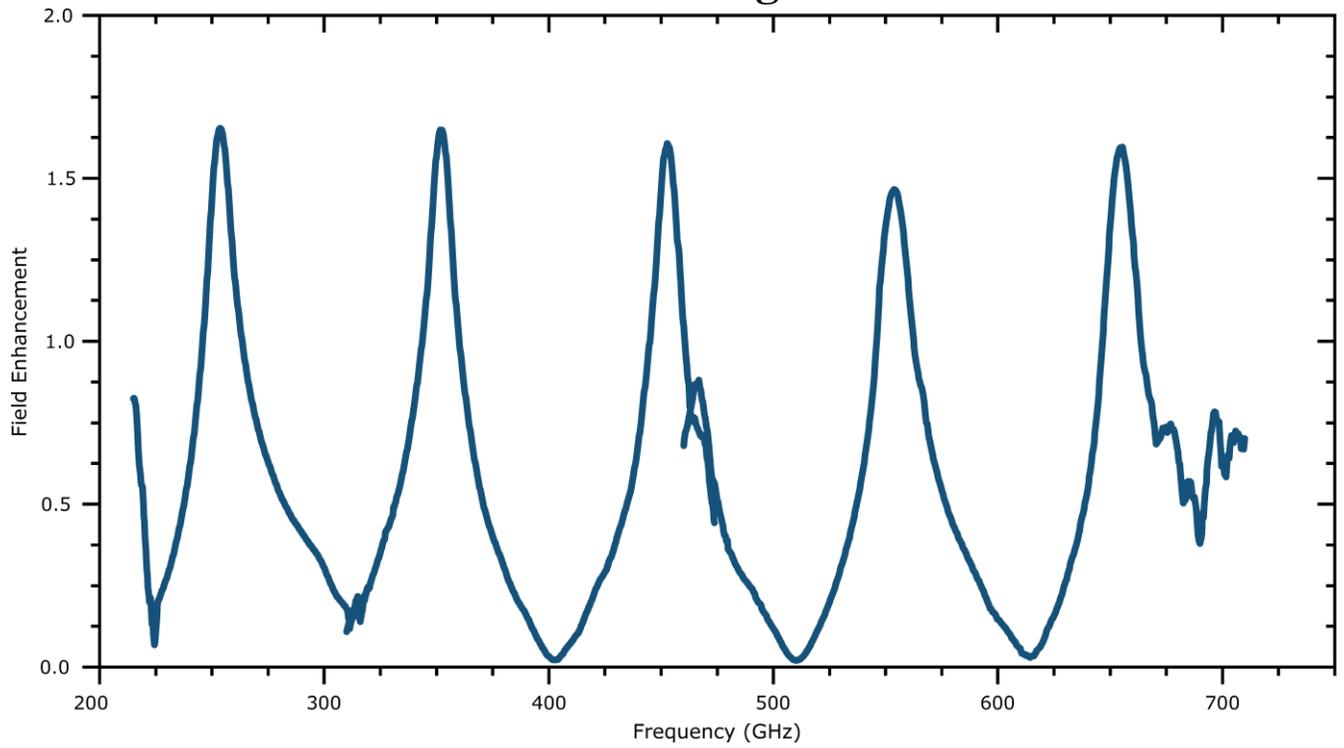

**Extended Data Fig. 1 | Field enhancement at the GaAs epilayer from the ITO-coated sapphire substrate.** The field enhancement is calculated as $|1 + r(f_{THz})|$ with the complex reflection coefficient $r(f_{THz})$ measured by a Vector Network Analyzer.

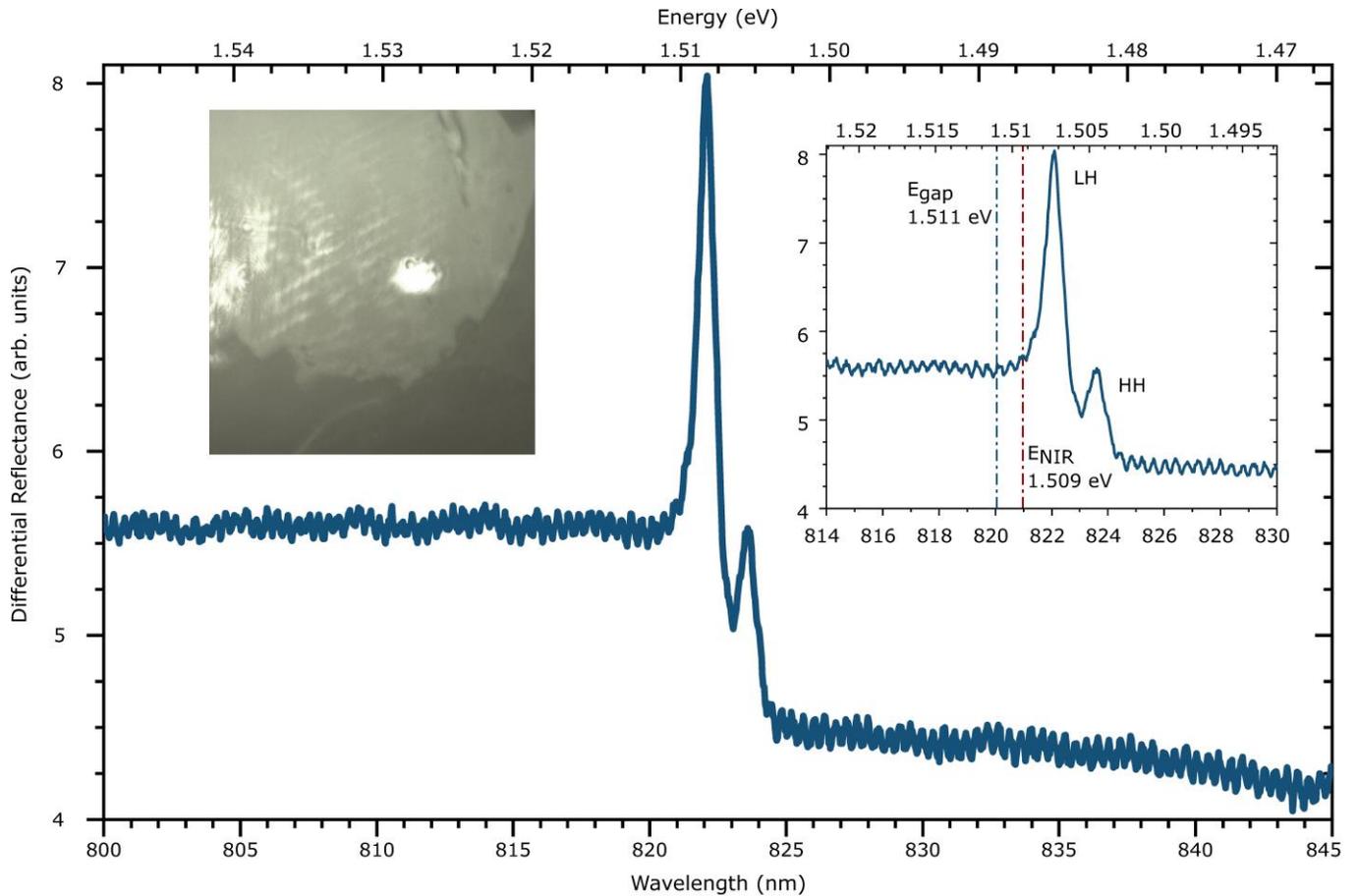

**Extended Data Fig. 2 | An absorbance spectrum of the GaAs epilayer mounted on the ITO-coated sapphire substrate.** The measurement was taken at the spot illuminated by a white light source (left inset). The right inset shows a zoom-in of the spectrum, with the bandgap and the photon energy of the NIR laser denoted by dash-dot blue and red lines, respectively. The two peaks are strain-split exciton resonances associated with band-edge states with different angular momenta. The temperature was 60K.

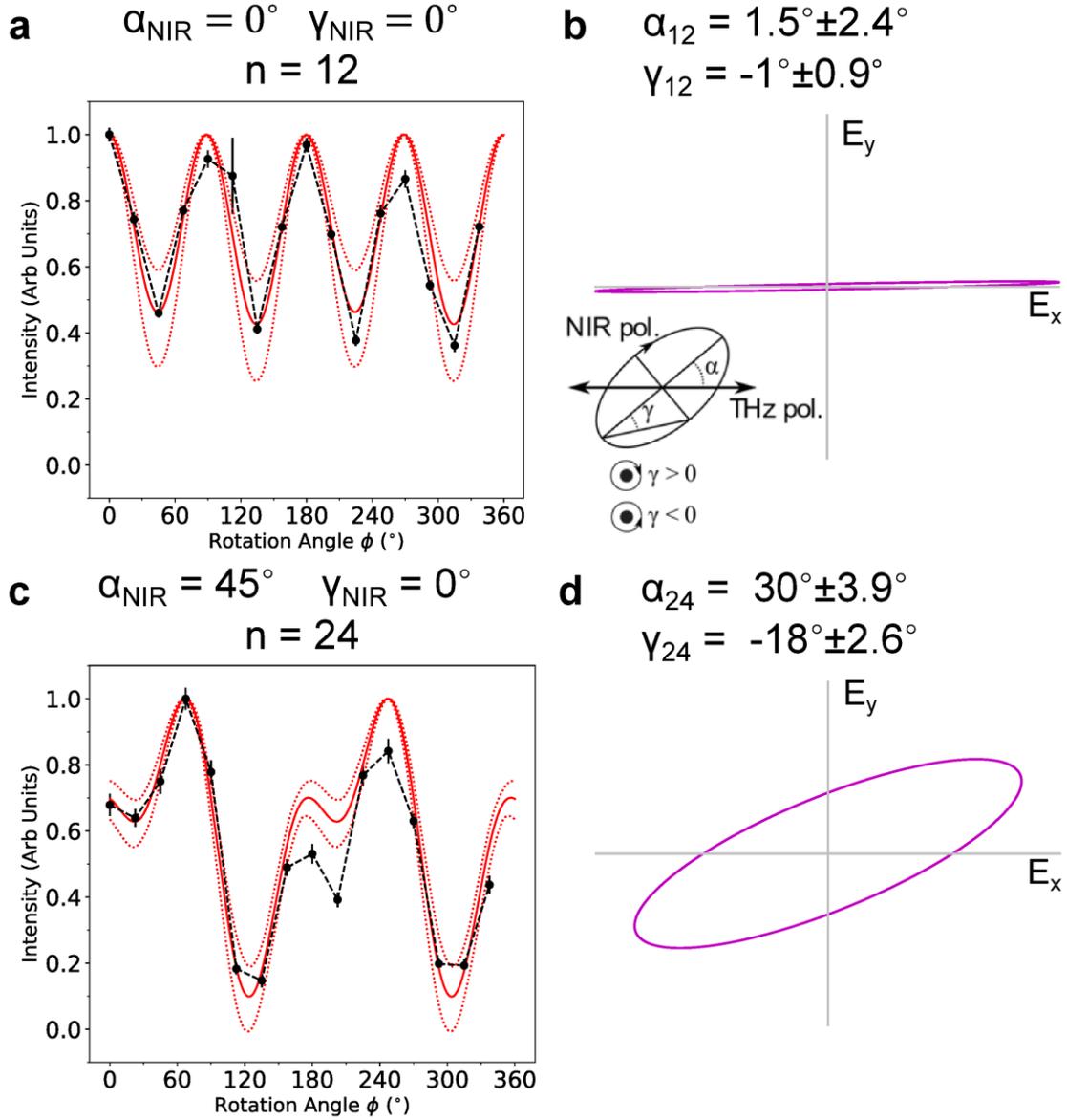

**Extended Data Fig. 3 | Stokes polarimetry with linearly polarized NIR laser ($\gamma_{\text{NIR}} = 0°$). a**, Polaragrams for sideband index $n = 12$ and orientation angle of the NIR laser $\alpha_{\text{NIR}} = 0°$. **b**, The polarization state of the sideband extracted from the polaragrams in **a**. **c**, Polaragrams for sideband index $n = 24$ and orientation angle of the NIR laser $\alpha_{\text{NIR}} = 45°$. **d**, The polarization state of the sideband extracted from the polaragrams in **c**. In **a** and **c**, the black dots show the measured polaragrams, with error bars showing the standard deviation over 4 measurements, and the red solid lines are the reconstructed polaragram through Fourier transform, with the red dotted lines showing the bounds. In **b** and **d**, the polarization states of the sidebands are represented as trajectories of the tips of the electric field vectors $(E_x, E_y)$ over time. The orientation angle $\alpha$ and ellipticity angle $\gamma$ are defined in the inset in **b**.

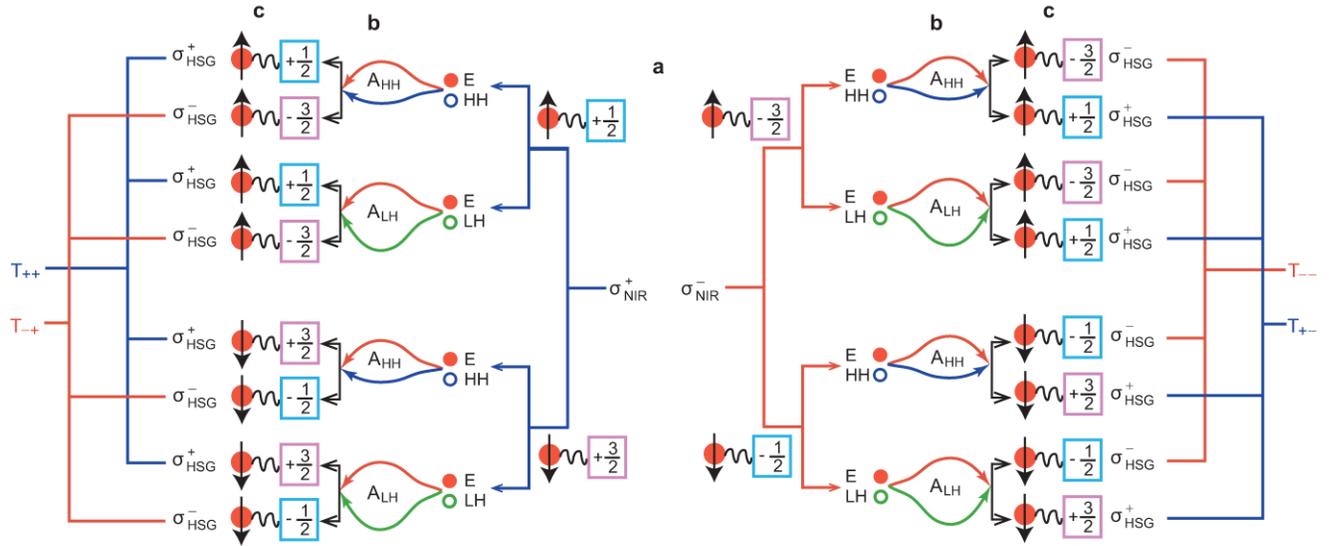

**Extended Data Fig. 4 | Quantum interference in three-step model of HSG leading to sideband polarization.** A photon from the NIR laser is decomposed into components $\sigma_{NIR}^{\pm}$, with helicity $\pm 1$. **a**, A $\sigma_{NIR}^{-}$ photon excites either a spin-up electron and hole of spin -3/2 or a spin-down electron and hole of spin -1/2. A $\sigma_{NIR}^{+}$ photon excites either a spin-up electron and hole of spin +1/2 or a spin-down electron and hole of spin +3/2. **b**, Driven by the THz field, an electron-hole pair accumulates dynamic phase $A_{HH}$ or $A_{LH}$, depending on the band of the hole state (HH or LH). The electron spin is unchanged, while the hole states originating from the spin -3/2 state are superpositions of spin -3/2 and +1/2 states and the states originating from the spin -1/2 state are superpositions of spin -1/2 and +3/2 states. **c**, Upon recollision, either $\sigma_{HSG}^{+}$ or $\sigma_{HSG}^{-}$ photons are produced following angular momentum conservation—for example, a spin +3/2 hole recombining with a spin-down (-1/2) electron produces a $\sigma_{HSG}^{+}$ photon with helicity +3/2 -1/2=+1. The interference of the evolution pathways from $\sigma_{NIR}^{\pm}$ to $\sigma_{HSG}^{+}$ ($\sigma_{HSG}^{-}$) produces the dynamical Jones matrix element $T_{+\pm}$ ($T_{-\pm}$).

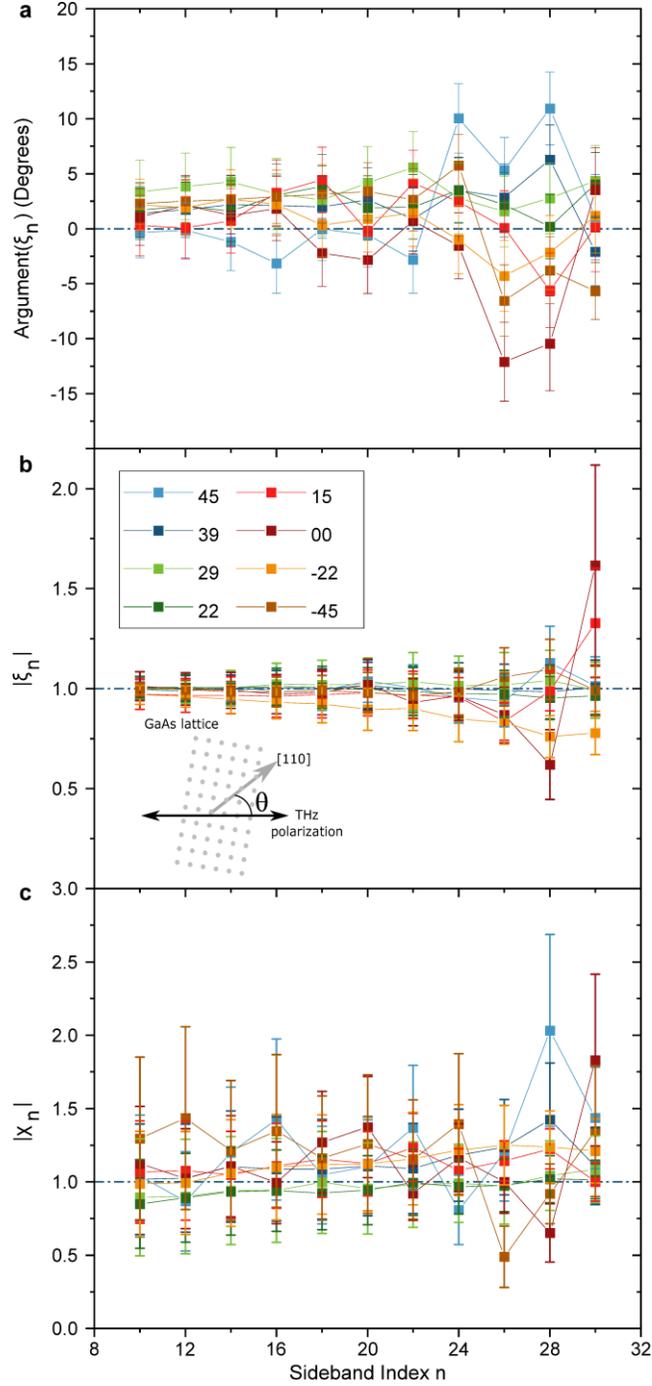

**Extended Data Fig. 5 | Additional data for ratios of Jones matrix elements, $\xi_n(\theta)$ and $\chi_n(\theta)$.** **a,** The argument of $\xi_n(\theta)$. The dash-dot line marks the expected value of 0. **b,** The magnitude of $\xi_n(\theta)$. The dash-dot line marks the expected value of 1. **c,** The magnitude of $\chi_n(\theta)$. The dash-dot line marks the expected value of 1. All quantities are presented as functions of sideband index $n$ for eight values of angle $\theta$. **Inset,** The definition of $\theta$ by using the GaAs crystal lattice and the THz electric field.

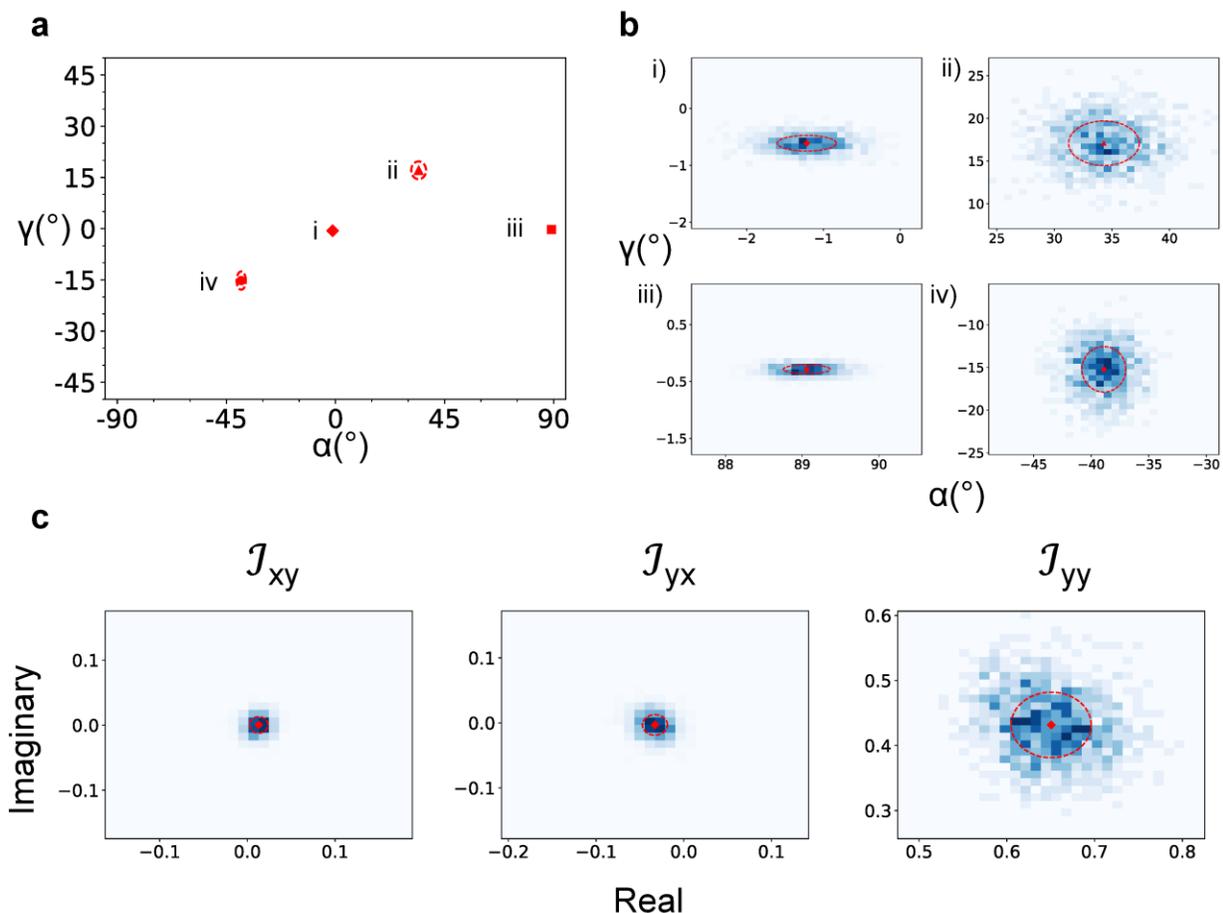

**Extended Data Fig. 6 | Monte Carlo simulation in calculating the dynamical Jones matrices.**
**a,** The polarization state of the $n = 12$ sideband ($\theta = 23°$) for all 4 initial NIR polarizations (i-$\alpha_{NIR} = 0°$, ii-$\alpha_{NIR} = 45°$, iii-$\alpha_{NIR} = 90°$, iv-$\alpha_{NIR} = -45°$). The horizontal and vertical axes represent $\alpha$ and $\gamma$, respectively. Dashed ovals correspond to confidence intervals in the measurement of $\alpha$ and $\gamma$. **b,** Histograms of $\alpha$ and $\gamma$ for the 4 measured sidebands polarizations. Normal distributions of $\alpha$ and $\gamma$ were sampled, with the central value and standard deviation of the distributions set by the measured values. In this figure, 1,000 iterations are shown, but the results of this paper are calculated from 10,000 iterations. **c,** The complex $\mathcal{J}$-matrix elements resulting from the $\alpha$ and $\gamma$ in **b**. The horizontal and vertical axes represent the real and imaginary part, respectively. Each red dashed line shows one standard deviation of the distribution of each $\mathcal{J}$-matrix element resulting from the Monte Carlo simulation. All three plots have the same scale. The value of $\mathcal{J}_{xx,n}$ is set as 1 in these calculations.

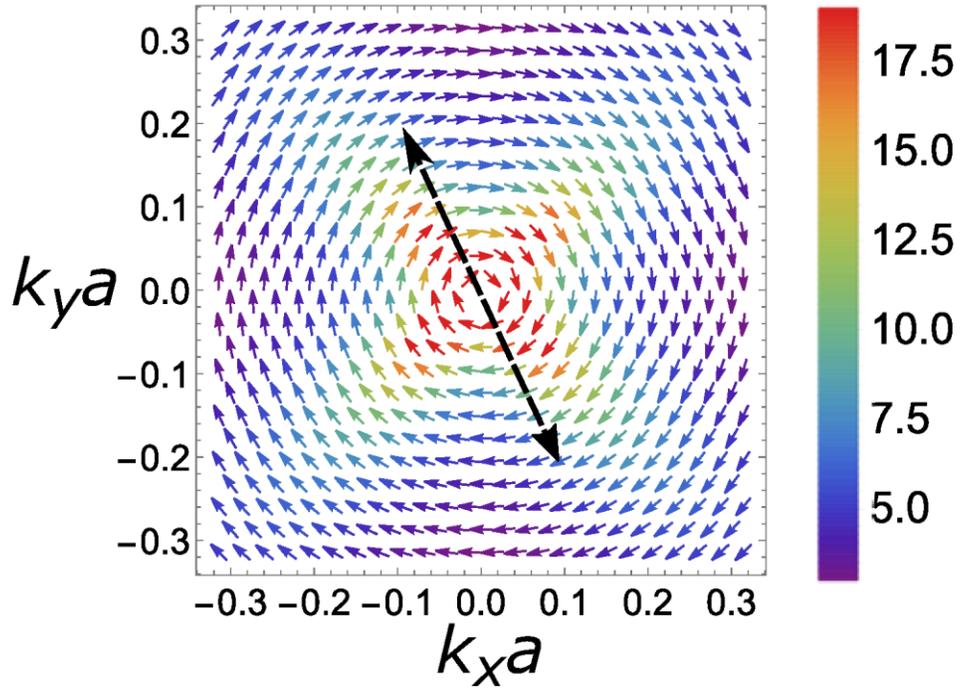

**Extended Data Fig. 7 | Berry connection matrix element $\mathcal{A}_{HH_+,HH_+}$ in the $k_z = 0$ plane of the Brillouin zone.** The double-headed black dotted arrow represents a path of a hole accelerated by a linearly polarized THz field, which is perpendicular to the Berry connection (color arrows) at all points. The Berry connection is plotted in units of $a$, which is the lattice constant of GaAs.

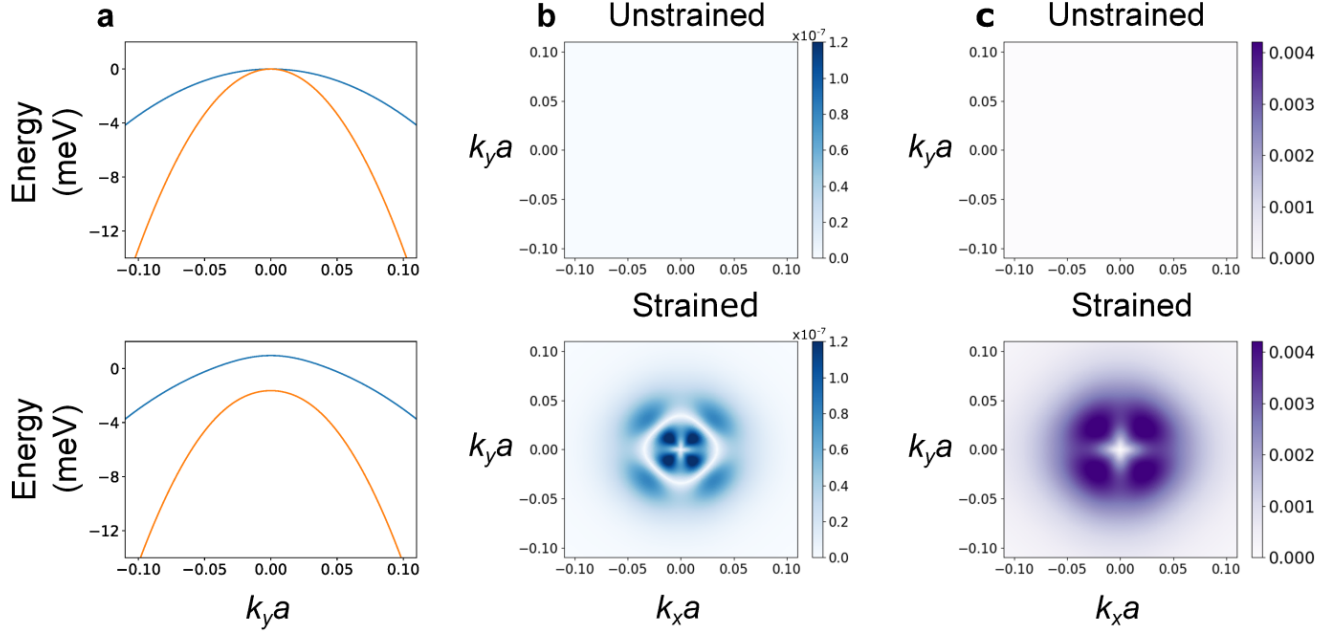

**Extended Data Fig. 8 | Effect of a biaxial strain on the valence band structure and non-Abelian Berry connection along the direction of quasimomentum $k$ in the $k_z = 0$ plane of the Brillouin zone.** The strain is chosen as tensile along [001] direction to be consistent with the splitting of the exciton peaks in the absorbance spectrum (Extended Data Fig. 2). **a,** Valence band structures along $k_x = k_z = 0$ for unstrained (top) and strained (bottom) GaAs. The blue and orange curves represent the heavy-hole and light-hole bands, respectively. **b,** The magnitude of the diagonal Berry connection matrix element $\mathcal{A}_{HH_+,HH_+}$ along the direction of quasimomentum for unstrained (top) and strained (bottom) GaAs. **c,** The magnitude of the off-diagonal Berry connection matrix element $\mathcal{A}_{HH_+,LH_+}$ along the direction of quasimomentum for unstrained (top) and strained (bottom) GaAs. For the unstrained case, the Berry connection along the quasimomentum is identically zero in the plots except for the singularity at $k=0$. The Berry connection is plotted in units of $a$, which is the lattice constant of GaAs.

27